%% file: main.tex
\documentclass[sigconf,nonacm]{acmart}
\input{package}

\begin{document}
\title{HoneyWin: High-Interaction Windows Honeypot in Enterprise Environment}
\author{Yan Lin Aung}
\authornote{This work was done while the author was at iTrust, Singapore University of Technology and Design.}
\email{y.aung@derby.ac.uk}
\affiliation{%
  \institution{University of Derby}
  \country{Derby, United Kingdom}
}
\author{Yee Loon Khoo}
\author{Davis Yang Zheng}
\authornote{Both authors contributed equally.}
\author{Bryan Swee Duo}
\authornotemark[2]
\affiliation{%
  \institution{Singapore Institute of Technology}
  \country{Singapore}
}
\author{Sudipta Chattopadhyay}
\author{Jianying Zhou}
\affiliation{%
  \institution{Singapore University of Technology and Design}  
  \country{Singapore}}
\author{Liming Lu}
\author{Weihan Goh}
\affiliation{%
  \institution{Singapore Institute of Technology}  
  \country{Singapore}}
\begin{abstract}
Windows operating systems (OS) are ubiquitous in enterprise Information Technology (IT) and operational technology (OT) environments.
Due to their widespread adoption and known vulnerabilities, they are often the primary targets of malware and ransomware attacks.
With 93\% of the ransomware targeting Windows-based systems, there is an urgent need for advanced defensive mechanisms to detect, analyze, and mitigate threats effectively.
In this paper, we propose \honeywin\ a high-interaction Windows honeypot that mimics an enterprise IT environment.
The \honeywin\ consists of three Windows 11 endpoints and an enterprise-grade gateway provisioned with comprehensive network traffic capturing, host-based logging, deceptive tokens, endpoint security and real-time alerts capabilities.
The \honeywin\ has been deployed live in the wild for 34 days and receives more than 5.79 million unsolicited connections, 1.24 million login attempts, 5 and 354 successful logins via remote desktop protocol (RDP) and secure shell (SSH) respectively.
The adversary interacted with the deceptive token in one of the RDP sessions and exploited the public-facing endpoint to initiate the Simple Mail Transfer Protocol (SMTP) brute-force bot attack via SSH sessions. 
The adversary successfully harvested 1,250 SMTP credentials after attempting 151,179 credentials during the attack.
\end{abstract}
\maketitle
\pagestyle{plain}
\input intro
\input honeywin
\input validation
\input results
\input discussion
\input related_work
\input conclusions
\section*{Acknowledgements}
This research is supported by the National Research Foundation, Singapore, under its National Satellite of Excellence Programme ``Design Science and Technology for Secure Critical Infrastructure: Phase II'' (Award No: NRF-NCR25-NSOE05-0001).
Any opinions, findings and conclusions or recommendations expressed in this material are those of the author(s) and do not reflect the views of National Research Foundation, Singapore.
\bibliographystyle{ACM-Reference-Format}
\bibliography{references}
\end{document}

%% file: package.tex
\usepackage[inkscapelatex=false]{svg}
\usepackage{pifont}

\usepackage{pgf-pie}
\usepackage{graphicx}
\usepackage{subcaption}
\usepackage[flushleft]{threeparttable}
\usepackage{ragged2e}
\usepackage{tabularx}
\usepackage{multirow}

\def\honeywin{\textbf{HoneyWin}}

\def\skipnoindent{\vskip0.1in\noindent}

\usepackage{algorithm}
\usepackage{algpseudocode}

%% file: intro.tex
\section{Introduction}
Windows operating systems (OS) are extensively utilized in enterprise Information Technology (IT) and operational technology (OT) environments, while their prevalence and roles differ across these domains.
In 2025, Windows OS maintains a dominant position holding more than 70\% of the global desktop OS market share, underscoring its widespread adoption in the enterprise IT environment~\cite{desktopshare:online}.
Similarly, Windows Server OS takes up a substantial market share of approximately 65\% in 2022~\cite{servershare:online}.
On the other hand, the use of Windows OS in OT environments, which encompass industrial control systems (ICS) and critical infrastructure, is also significant.
Many OT systems, such as Supervisory Control and Data Acquisition (SCADA) and Human-Machine Interface (HMI) applications, run on Windows-based platforms.
The reference architecture developed by ATT\&CK sets out Windows-based systems that are used as application server, engineering workstation, transient cyber asset (TCA), safety engineering workstation, etc. in OT environments~\cite{refarch:online}.
Due to compatibility requirements with existing industrial hardware and software, several OT environments continue to operate on legacy Windows systems, such as Windows XP and Windows 7.

A recent incident with the CrowdStrike software update highlights a critical dependency on Windows-based systems.
8.5 million devices are affected, causing widespread disruptions in multiple industries worldwide~\cite{crowdstrike:online}.
Due to their widespread use and known vulnerabilities, Windows systems are often the primary targets of malware and ransomware attacks.
In addition, Windows systems in OT environments may not always be configured with stringent security measures, potentially leading to unauthorized access and system compromises~\cite{winot:online}.
Reliance on legacy Windows systems in OT settings poses significant security challenges and requires proactive measures to mitigate associated risks.

The rise in ransomware attacks over the past few years further underscores the urgent need for advanced defensive mechanisms.
With 93\% of the ransomware targets Windows-based systems, enterprises require robust countermeasures to detect, analyze, and mitigate threats effectively~\cite{ransomwarestats:online}.
Typically, honeypot systems are deployed by enterprise security teams, IT departments, government agencies, threat intelligence companies, cloud service providers, security researchers, and academics as an effective cyber security strategy for this purpose.
They provide valuable threat intelligence, improve defensive capabilities, and serve as an early warning system against cyber threats.
A variety of honeypot systems have been proposed and developed.
Notable implementations include Cowrie~\cite{cowire:online}, Glutton~\cite{glutton:online}, OpenCanary~\cite{opencanary:online}, Conpot~\cite{conpot:online}, T-Pot~\cite{tpot:online}, AIDE~\cite{aide:online}, ICSNet~\cite{icsnet}, SIPHON~\cite{guarnizo2017siphon} etc.
However, despite previous work such as~\cite{honey:monkeys,kfsensor:online}, there has been very little state-of-the-art research, HopLab in~\cite{hoplab:online} for instance, on Windows-based honeypot systems, which has created a significant research gap.
On the other hand, conventional security solutions focus primarily on detection and response.
However, a proactive approach involving deception could significantly enhance an organization's security posture~\cite{sanswhitepaper:online}.
The need for additional security layers that do not depend solely on conventional endpoint protection solutions has never been more apparent~\cite{endpoint:online}.

To address these concerns, this paper introduces \honeywin\, a high-interaction Windows honeypot that mimics an enterprise IT environment.
The \honeywin\ incorporates Windows 11 endpoints and an enterprise-grade firewall provisioning with network traffic capture, host-based event logging, deceptive tokens, endpoint security (EDR) and real-time intrusion detection alerts capabilities.
Based on our review of existing state-of-the-art works, \honeywin\ is a first-of-its-kind high-interaction Windows honeypot that has been designed, implemented, validated and deployed live in the wild.

\honeywin\ offers the following contributions:
\begin{itemize}
    \item \textbf{A scalable high-interaction Windows honeypot design.}
    The proposed \honeywin\ system consists of three Windows 11 endpoints and an enterprise-grade firewall.
    However, \honeywin\ was designed for scalability in the first place to allow us to conveniently expand it in the future.
    In particular, \honeywin\ incorporates a lightweight container-based approach to establish public-facing honeypot devices.
    Both incoming and outgoing network traffic is captured separately with dedicated systems, not within the honeypot devices.
    Host event logs are not stored within the honeypot devices; instead, they are shipped directly to avoid detection and tamper proofing.
    \item \textbf{Holistic detection capability.}
    The \honeywin\ captures both network traffic and host event logs.
    The honeypot devices are equipped with a commercial state-of-the-art endpoint security solution.
    By this means, \honeywin\ accommodates the correlation between network traffic and host event logs and facilitates a holistic detection capability.
    \item \textbf{Deceptive tokens implementation and deployment.}
    To mislead and deceive adversaries who gain access to the honeypot devices, the \honeywin\ implementation incorporates two types of deceptive tokens: (1) Spoofed Windows commands -- commonly used Windows discovery commands that have been reconfigured to deceive attackers into the realm of navigating a legitimate enterprise network environment and (2) realistic bait files to further enhance the deception.
\end{itemize}
\textbf{Organization:} The remainder of this paper is organized as follows.
Section~\ref{sec:honeywin} introduces \honeywin\ beginning with a threat model and various design considerations, followed by the implementation of \honeywin\ in an enterprise environment.
Section~\ref{sec:validation} discusses penetration testing and malware detection testing to validate the implementation of \honeywin.
We provide experimental results from live deployment of the \honeywin\ system in the wild for 34 days, including an in-depth analysis on the attacks received.
Section~\ref{sec:discussion} discusses the insights from the design and implementation of the \honeywin\ system and the experimental results.
Finally, we provide related work in Section~\ref{sec:related-work} and conclude the paper in Section~\ref{sec:conclusions}.
%
%
%

%% file: honeywin.tex
%
\section{HoneyWin: High Interaction Windows Honeypot in an Enterprise Environment} \label{sec:honeywin}
This section describes the proposed \honeywin\ system.
Firstly, we present the threat model and considerations that have been taken into account when designing the \honeywin.
Subsequently, we provide details of the implementation of \honeywin.
\subsection{Threat Model} \label{subsec:threat-model}
We assume that the attacker has access to the honeypot by scanning the Internet or using a search engine such as Shodan\footnote{\scriptsize \url{https://www.shodan.io/}}.
Once the honeypot is selected as the target, the attacker initiates a reconnaissance using port scan tools such as \texttt{nmap}\footnote{\scriptsize \url{https://nmap.org/}} and identifies the exposed services (e.g., RDP port 3389).
The attacker then probes each service to obtain more information and attempts to authenticate with a username and password.
In this case, the attacker may brute-force or use other possible means to have the correct credentials.
Upon gaining access, the attacker advances with the next phase of the cyber kill chain.
The honeypot is designed to log the attacker's interactions via network traffic and host event logs.
In addition, the honeypot is implemented with a real-time alerting mechanism to report critical events, such as successful logins.
\subsection{Design} \label{subsec:design}
This section discusses various considerations taken into account when designing a Windows-based honeypot in an enterprise environment.

\textbf{High-Interaction Honeypot with Real Systems:}
We anticipate that the proposed honeypot setup includes real systems instead of low-interaction implementation that emulated certain services (e.g., Cowrie SSH/Telnet honeypot~\cite{cowire:online}).
Having real systems as high-interaction honeypots maximizes the attack surface and allows full access to the underlying system.
The setup features Windows 11 endpoints that are accessible directly on the Internet.
Moreover, the setup incorporates an enterprise-grade gateway/firewall with Windows endpoints connected on the local area network (LAN), while the wide area network (WAN) side of the gateway is exposed on the Internet mimicking a typical enterprise environment.

\textbf{Exposing Honeypots on the Internet:}
It is imperative that Windows endpoints and the enterprise gateway require public IP addresses to make them accessible on the Internet.
Windows endpoints and the gateway could be hosted with Infrastructure as a Service (IaaS) or Virtual Private Server (VPS) which belong to cloud service providers (e.g. Amazon Web Services).
Using Virtual Private Network (VPN) service provides public IP addresses from servers located in more diverse geolocations while the devices could remain within the perimeter of our setup.

\textbf{Network Traffic Collection and Host-based Event Logging:}
To detect and analyze intrusions and threats received by devices in our honeypot, it is essential to capture all incoming network traffic.
In addition, we must capture the outgoing network traffic initiated from the devices, as these may be attempts by malicious actors to establish network connections to command \& control (C2) servers, scanning and targeting other vulnerable devices elsewhere.
Modern operating systems (e.g., Windows, Linux) and security appliances (e.g., Cisco, Fortinet) feature logging of events related to the system, security, and applications.
Host logs provide fine-grained visibility into the activities that occur in each device, allowing better detection, investigation, and response that complements network traffic analysis.

\textbf{Incorporation of Deceptive Tokens:}
Deceptive tokens are crucial to making the honeypot a realistic high-interaction environment providing a high level of interactivity with the system while preventing attackers from actually affecting the system and network.
The deceptive tokens in the form of windows built-in executables, would provide benefits such as being easily configurable, consistent, and scalable across multiple honeypot deployments.
The spoofed windows commands reading from a common configuration file that is easily replicable allow for quick deployment across multiple honeypot devices. 
In addition, a modular design approach allows partial modifications without affecting the entire configuration and making updates seamless.
To deceive the attackers, these commands would be reconfigured so that they deceive the attackers into believing that they are navigating a real corporate network environment.
In the event that an attacker gains access via remote means, deceptive tokens ensure that the system presents misleading but realistic network and system information.
By designing the tokens in the form of built-in system commands, these modified executables generate and return false, yet plausible system responses, reinforcing the illusion of a convincing corporate network.
This approach enhances adversary engagement while allowing defenders to gather valuable intelligence on the behavior of the attacker.

\textbf{Real-time Intrusion Detection and Alerts:}
Having Windows endpoints and the gateway accessible from anywhere implies a rich attack surface.
Moreover, there is a significant security risk once the attacker compromises and takes control of the systems.
To mitigate such risk and take appropriate actions, the setup should implement real-time intrusion detection (e.g., successful logins) and alerting mechanism.

\textbf{Network Traffic and Log Backup:}
Through the reconnaissance, the attacker gains access to the Windows endpoints and gateway.
Therefore, network traffic collection shall not take place within the systems in the first place.
Similarly, the host logs shall not be kept within the systems; instead, they should be forwarded and stored in a security-hardened system.
Even in such a scenario, it is still possible that either the collected network traffic or host logs are compromised.
To minimize the risk of losing such invaluable data, the set-up shall incorporate a backup system to store the replica of captured data and logs.
The backup system shall not have an Internet connection and has access to the systems storing the data and logs but not the other way around.

\textbf{Cloning and Restoring Windows Systems:}
It is necessary to save the system image before they are deployed live on the Internet since it is likely that the systems get compromised over time and may become out of control.
Once sufficient data are captured to attribute the attack tactics, techniques and procedures (TTPs), the systems could be restored into their pre-deployment state.
Depending on the attack vectors, certain adjustments to the system image may be necessary before redeploying live on the Internet.

\textbf{Log Management and Analysis Platform:}
To efficiently manage and analyze large volumes of log data, it is essential to incorporate a log management and analysis platform such as ELK\footnote{\scriptsize \url{https://www.elastic.co/elastic-stack}} or Splunk\footnote{\scriptsize \url{https://www.splunk.com/}} into our setup.
These platforms are capable of collecting, parsing and storing large amounts of data, as well as searching, filtering, analyzing, and visualizing the collected data.

\textbf{Automation and Orchestration:}
It is expected that the honeypot is deployed live around the clock.
Hence, various design considerations discussed above such as setting up of VPN connections, network traffic and host logs collection, intrusion detection and alerts, etc. shall be automated.
In addition, monitoring of the status of the entire setup and orchestration of various components independently are required.
\subsection{Implementation}
\label{subsec:implementation}
%
%
Taking into account the design considerations discussed in Section~\ref{subsec:design}, we envision and propose high-interaction Windows honeypot system in an enterprise environment, namely \honeywin.
Figure~\ref{fig-honeywin} shows the proposed implementation of \honeywin.
\input{figures/fig-honeywin}
The setup consists of three Windows 11 endpoints (\texttt{E1}, \texttt{E2} and \texttt{E3}) and an enterprise gateway (\texttt{G1}) as honeypots.
The endpoint \texttt{E3} and the gateway \texttt{G1} are directly accessible on the Internet, while the endpoints \texttt{E1} and \texttt{E2} reside within the private network (i.e., the `Enterprise Network' (EPN) in Figure~\ref{fig-honeywin}).
They are connected to the gateway and have Internet connection via Network Address Translation (NAT) scheme.
The incoming connection therefore must bypass the gateway to access the \texttt{E1} and \texttt{E2} endpoints.

\textbf{VPN Forwarding with Docker Container:}
To expose devices on the Internet, we adopted a lightweight approach to establish secure tunnels to VPN servers and acquire public IP addresses.
In particular, a customized docker container has been developed with the functionalities to establish VPN tunnels to servers in specified geolocations, set up port forwarding to the devices on the `Honeypot Network' and automatically capture the incoming network traffic received on exposed public IP addresses.
Typically, each device (i.e., \texttt{E3} and \texttt{G1}) has a dedicated docker container.
However, it should be noted that our implementation allows multiple containers to forward the network traffic to the same device, thereby increasing the geographic presence given a limited number of real systems in case.
The containers run inside the Workstation (\texttt{U0}).
To automate the number of containers and their names, public IP addresses, open ports, and forwarding IP addresses to the devices on the `Honeypot Network' (HPN), a set of shell scripts and a CSV file are used.
To expose a specific device on the Internet, the user first connects the device to the HPN, then updates the CSV file with the necessary information, and uses the script to establish end-to-end connectivity between the VPN server and the device.
With the aid of a shell script and also thanks to a docker container, we could bring all containers or individual ones online or offline within seconds.
In addition, the IP addresses of the exposed devices are mapped to a registered domain to mimic an enterprise environment.

\textbf{Incoming and Outgoing Network Traffic Collection:}
As described in the previous section, incoming network traffic to public-facing devices is captured in each docker container.
Once the attacker is within the HPN or the EPN, outbound connections may be initiated to contact C2 servers, scan and target other vulnerable devices elsewhere since all devices on both networks have the Internet connection.
We use the SPAN port of the switch (\texttt{S1}) to capture the outgoing network traffic and store the PCAPs in the Workstation (\texttt{U1}).

\textbf{Host Log Collection and Monitoring with ELK Stack:}
We also collect host logs from the honeypot devices.
For all Windows endpoints, the System Monitor (Sysmon) is installed to monitor and log system activity to the Windows event log.
Sysmon\footnote{\scriptsize \url{https://learn.microsoft.com/en-us/sysinternals/downloads/sysmon}} provides detailed information on process creations, network connections, and changes to file creation time.
Sysmon runs as a protected process and does not allow for a wide range of user-mode interactions.
However, Sysmon does not provide analysis of the events and does not attempt to hide itself from attackers.
As such, we rename the Sysmon executable and the driver to hide it and add complexity to adversaries attempting to identify security tools in use as a first line of defense.
For the enterprise gateway, we enable Syslog logging via the administrator dashboard.

Instead of storing the host logs within the honeypot devices, we have setup an ELK stack in Workstation (\texttt{U1}) for log management and analysis.
In this case, an Elastic Agent (EA) is installed in each Windows-based system.
For the gateway (\texttt{G1}), an intermediate system is setup and the Elastic Agent is installed to collect and ship the Syslog to \texttt{U1}.
The Elastic Agent\footnote{\scriptsize \url{https://www.elastic.co/elastic-agent}} provides a unified way to add monitoring for logs, metrics, and other types of data to a host device.
To prevent the EA from being uninstalled without authorization, agent tamper protection is also enabled in the EA policy.
We have also set up Fleet\footnote{\scriptsize \url{https://www.elastic.co/guide/en/fleet/current/fleet-overview.html}} in the ELK stack and installed the Fleet Server in \texttt{U1} to manage EAs and their policies.
Elastic provides integrations, which are prepackaged assets allowing users to collect, store, and visualize data from various sources with ELK.
Elastic Defend\footnote{\scriptsize \url{https://www.elastic.co/guide/en/integrations/current/endpoint.html}} is one of the available integrations in ELK to detect, prevent, and respond to cyber threats.
We have installed Elastic Defend, Prebuilt Security Detection Rules, Windows, and enterprise gateway logs integrations with our setup.
Prebuilt Security Detection Rules integration stores the security rules to detect malware in the endpoints and generate alerts.
Windows integration is to receive the Sysmon logs forwarded by the endpoint EA whereas enterprise gateway logs integration is to receive the Syslog from the intermediate system of the gateway, respectively.
\input{tables/tab-deceptive-tokens}

\textbf{Successful Login Alerts:}
Based on our threat model in Section~\ref{subsec:threat-model}, the attacker attempts to log in to our honeypot devices through the exposed services.
In our setup, we have opened RDP (i.e. Port 3389) and SSH (i.e. Port 22) for Windows endpoints, whereas HTTPS (i.e. Port 443) is open for access to the administrator dashboard of the enterprise gateway.
The attacker who is able to successfully log in with correct credentials signifies a breach to our devices.
Hence, it is critical to provide real-time notifications of successful login to any of our devices.
To realize this capability, we have installed Winlogbeat in our Windows endpoints and configured it to capture successful RDP and SSH login events, and forward the event logs to the Logstash component installed in \texttt{U1}.
We then use a webhook integration to send out emails to notify the successful login events.

In addition, we have also configured the Winlogbeat to capture failed logins, which indicate brute-force attempts.
The events are forwarded and stored in \texttt{U1}.
For the case of enterprise gateway, we have created an automation with `Admin login successful' event as a trigger.
Upon activation, the webhook sends an email alert.

\textbf{User Accounts and Login Credentials:}
For each Windows endpoint, we have one user account with administrative privileges and one or more standard users.
The password for the administrative user is a randomly generated one with 16 characters consisting of uppercase, lowercase, digits, and brackets.
This has been the design decision to prevent the adversary from taking complete control of the endpoint.
However, the passwords for standard users are weak passwords with just eight characters consisting of uppercase, lowercase, and digits.
For the gateway (\texttt{G1}), there is only one administrative user and the password is set as a random generated one with 16 characters consisting of uppercase, lowercase, digits, and brackets.

\textbf{Firewall and Enterprise Gateway:}
In Figure~\ref{fig-honeywin}, we have set up a `Firewall \& Router' to manage VLAN trunks, DHCP reservations, DNS settings, firewall rules, etc.
For example, the outgoing connections from the HPN are blocked by default.
Selected devices are given Internet access through firewall rules.
Similarly, the enterprise gateway/firewall (\texttt{G1}) is also configured to have fine-grained control and access management of devices in the EPN.

\textbf{Backup Server and Clonezilla:}
We have setup Workstation (\texttt{U2}) as a backup server to store the incoming network traffic captured in \texttt{U0}, the outgoing network traffic captured via the SPAN port with \texttt{U1}, successful logins and failed login attempts, and Elasticsearch indices of Sysmon events.
The \texttt{U2} connects to \texttt{U0} and \texttt{U1} on the `Backup Network' (BKN) and does not have the Internet connection.
Also, only \texttt{U2} has access to \texttt{U0} and \texttt{U1} but not vice versa.
We use Clonezilla to keep the system image of Windows endpoints so as to restore into the pre-deployment state if the system gets compromised and becomes out-of-control.
In addition, we save the configurations of `Firewall \& Router' and enterprise gateway (\texttt{G1}).

\textbf{Scalable Design:}
While the current \honeywin\ setup consists of a limited number of honeypot devices: an enterprise gateway and three Windows endpoints, we have designed it for scalability which enables us to expand it conveniently in the future.
Having a lightweight container-based approach to establish secure VPN tunnels, we could extend the setup with additional public-facing endpoints.
Moreover, we could place and interchange the honeypot devices at different geolocations during specific time periods.
Capturing of incoming and outgoing network traffic inside each container and U1 respectively, real-time alerting of successful logins to the honeypot devices, backing up of PCAPs and Elasticsearch indices are fully automated.
The state-of-the-art ELK stack with Elastic Defend, Prebuilt Security Detection Rules, Windows and Firewall Logs integrations, as well as Fleet and Elastic Agent allows us to secure the endpoints from tampering, detect, and analyze TTPs.

\textbf{Deceptive Tokens:}
We have also incorporated deceptive tokens into the Windows endpoints.
Table~\ref{tab-deceptive-tokens} shows the deceptive tokens allocation and installation in three Windows endpoints.
The objective is to mislead adversaries who gain access to the system by manipulating the output of commonly used Windows discovery commands, such as \texttt{systeminfo}, \texttt{ipconfig}, \texttt{netstat}, \texttt{net}.

\texttt{`systeminfo'} is commonly used for host enumeration by attackers as it provides detailed system information, useful for attackers to plan their next course of action, as key details include windows version, build number and installed patches along with domain membership and hostname as well as other useful data.
\texttt{`ipconfig'} is used for network reconnaissance as it provides critical networking information that could be used for lateral movement.
It could also be used to list Domain Name System (DNS) and Dynamic Host Configuration Protocol (DHCP) servers that could be further leveraged by attackers to perform other network based attacks such as DNS poisoning.
\texttt{`netstat'} is used for network reconnaissance as it lists active remote connections in the host system and provides a list of connected hosts that could be pivoted to by the attacker.
\texttt{`net'} allows adversaries to list user accounts on the local system as well as domain users and their group members, giving crucial data for privilege escalation on systems or allowing attackers to install a backdoor by using the binary to add or remove users from the system.
Additionally, it allows the enumeration of network shares for data exfiltration or lateral movement, making the executable highly valued for living off the land movement.
The deceptive tokens listed in Table~\ref{tab-deceptive-tokens} were specifically selected and created to replace built-in windows executables that have been identified to be key binaries exploited by adversaries/nation state actors during their operations where part of their unique trade craft involved.

After the design and creation of the deceptive tokens, implementation in the \honeywin system requires careful execution to prevent attackers from detecting legitimate binaries.
The deceptive tokens and bait are distributed in various locations on the different \honeywin\ systems, as provided in Table~\ref{tab-deceptive-tokens}.
\texttt{`ipconfig'}, \texttt{`net'}, \texttt{`netstat'}, \texttt{`arp'} and \texttt{`systeminfo'} were placed on all three hosts on \texttt{`C:\textbackslash Windows\textbackslash System32\textbackslash wbem'} together with \texttt{`wmiutil.mof'} which contains the configuration data used by the different binaries.
The \texttt{`wmiutils.mof'} file is placed in the Web-Based Enterprise Management folder to hide among other Managed Object Format files that are used by the Windows Management Instrumentation executable.
The altered binaries were placed there to enhance the legitimacy of existing within the \texttt{`System32'} folder and are added to the user path from there to take precedence over the original binaries, which are left in their original locations for future use.

To further enhance the deception, the team has deployed realistic bait files along with these deceptive executables.
These files include configuration files, project meeting minutes, network infrastructure reports, and AWS Identity and Access Management (IAM) credentials that appear to provide access to cloud services.
By planting such high-value artifacts, the system attracts attackers to interact with the deceptive environment, thereby increasing the likelihood of engagement and intelligence collection.
It should be noted in Table~\ref{tab-deceptive-tokens} that \texttt{<E2>} and \texttt{<E3>} are placeholders and represent standard users for the endpoints.
As mentioned earlier, each endpoint may have more than one standard user.

%% file: figures/fig-honeywin.tex
\begin{figure*}  
  \includegraphics[width=0.95\textwidth,keepaspectratio]{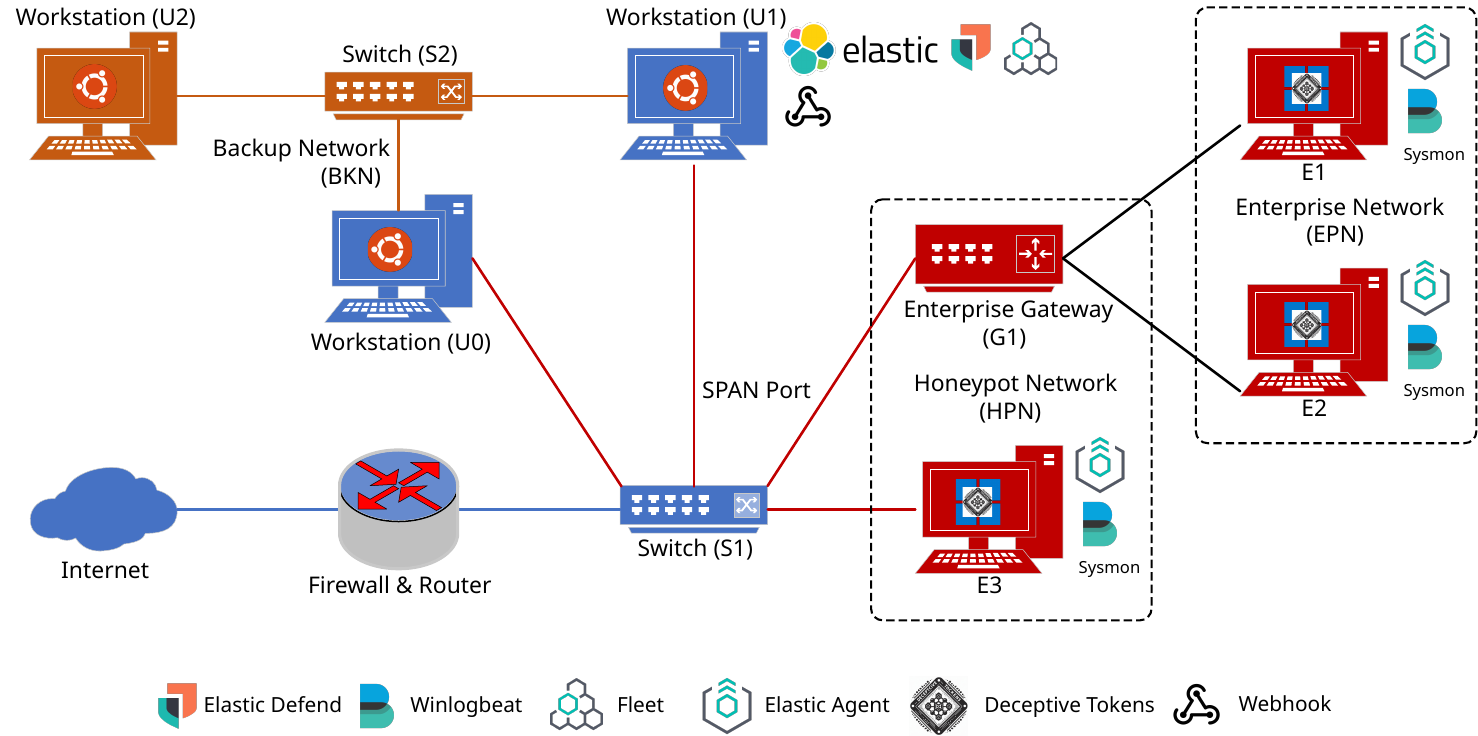}
  \caption{Windows-based Honeypots in Enterprise Environment} 
  \label{fig-honeywin}  
  \Description{}
\end{figure*}

%% file: tables/tab-deceptive-tokens.tex
\begin{table*}[!htbp]
    \centering
    \caption{Deceptive Tokens, Installation Paths and Hostnames}
    \label{tab-deceptive-tokens}
    \begin{threeparttable} 
    \begin{tabularx}{0.75\textwidth}
    {>{\hsize=1.65\hsize\linewidth=\hsize}X|>
    {\hsize=0.95\hsize\linewidth=\hsize}X|>    
    {\hsize=0.40\hsize\linewidth=\hsize}X
    }
        \toprule \toprule
        \multicolumn{1}{c|}{Deceptive Token} & \multicolumn{1}{c|}{Path} & \multicolumn{1}{c}{Hostname} \\
        \cline{1-3}
        \RaggedRight{\texttt{wmiutils.mof}} & \multirow{6}{*}{\RaggedRight{\texttt{C:\textbackslash Windows\textbackslash System32\textbackslash wbem}}} & \multirow{6}{*}{\RaggedLeft{\texttt{E1, E2, E3}}} \\
        \RaggedRight{\texttt{ipconfig.exe}} & & \\
        \RaggedRight{\texttt{net.exe}} & & \\
        \RaggedRight{\texttt{netstat.exe}} & & \\
        \RaggedRight{\texttt{arp.exe}} & & \\
        \RaggedRight{\texttt{systeminfo.exe}} & & \\
        \cline{1-3}
        \RaggedRight{\texttt{Current-5G-Network-Infrastructure-PPP-in-\newline H2020\_Final\_November\_2023.pdf}} & \multirow{4}{*}{\RaggedRight{\texttt{C:\textbackslash Users\textbackslash <E3>\textbackslash Desktop}}} & \multirow{6}{*}{\RaggedLeft{\texttt{E3}}} \\
        \RaggedRight{\texttt{5G\_vendor\_technology\_list.xlsx}} & & \\
        \RaggedRight{\texttt{TDoc\_List\_meeting\_SA\#91-e.xlsx}} & & \\
        \RaggedRight{\texttt{Security-in-5G.pdf}} & & \\
        \cline{1-2}
        \RaggedRight{\texttt{db\_info.conf}} & \multirow{2}{*}{\RaggedRight{\texttt{C:\textbackslash Users\textbackslash <E3>\textbackslash Documents}}} & \\
        \RaggedRight{\texttt{iam\_access.conf}} & & \\
        \cline{1-3}
        \RaggedRight{\texttt{aws\_creds-list\_14082024.xlsx}} &  \multirow{2}{*}{\RaggedRight{\texttt{C:\textbackslash Users\textbackslash <E2>\textbackslash Documents}}} & \multirow{5}{*}{\RaggedLeft{\texttt{E2}}} \\
        \RaggedRight{\texttt{intelsat-mobile-live-5G-infra.pdf}} & & \\
        \cline{1-2}
        \RaggedRight{\texttt{ESEC Structure.png}} & \multirow{3}{*}{\RaggedRight{\texttt{C:\textbackslash Users\textbackslash <E2>\textbackslash Desktop}}} \\
        \RaggedRight{\texttt{huawei-5g-cpe-pro-custom.pdf}} & & \\
        \RaggedRight{\texttt{IC-Project-Team-Meeting-Minutes-11856.xlsx}} & & \\
        \bottomrule \bottomrule
    \end{tabularx}
    \end{threeparttable}
\end{table*}

%% file: validation.tex
\section{Validation of HoneyWin} 
\label{sec:validation}
The primary objective of evaluating the \honeywin\ implementation is to confirm its effectiveness in mimicking a realistic enterprise environment and capturing real-world attack behaviors.
In particular, the validation process has been designed to measure how effectively the honeypot detects and responds to network reconnaissance, unauthorized access attempts, and privilege escalation, while also determining its overall detection visibility, response time, and ability to capture all intended data.

By simulating common adversarial tactics, ranging from port scans and lateral movements to sophisticated privilege escalation methods, the validation effort provided detailed insight into monitoring granularity, generating alerts, and the ability to withstand more advanced attack strategies targeting \honeywin.
This rigorous testing not only validates the design assumptions of \honeywin\ but also highlights areas for further refinement, ensuring that it provides high fidelity in both deception and detection capabilities.

\subsection{Penetration Testing}

Penetration testing formed the cornerstone of the validation process for \honeywin, ensuring that its network monitoring and logging systems are functional and collect the desired data.
Before incorporating the deceptive tokens, the environment is rigorously tested to confirm proper implementation.

\textbf{Network Reconnaissance:} A series of \texttt{`nmap'} scans are simulated adversarial port-scanning behavior.
These scans revealed limited exposure and strong initial defenses:
\begin{itemize}
    \item \textbf{Top 1000 and all available ports:} Port 514 was consistently displayed as filtered, demonstrating restrictive policies typical of secure environments.
    \item \textbf{Targeted scans on ports 22 (SSH) and 3389 (RDP):} Both returned “host down” statuses, underscoring effective blocking of unnecessary exposure.
\end{itemize}

\textbf{Host-Level Testing:} Tools and commands such as \texttt{`netstat'}, \texttt{`ipconfig'}, \texttt{`route print'}, and \texttt{`arp'} were executed to evaluate network configurations and logging capabilities of \honeywin.
The results demonstrated that logging configurations effectively track host-specific actions.

Windows Privilege Escalation Awesome Scripts (WinPEAS)~\footnote{\scriptsize \url{https://github.com/peass-ng/PEASS-ng/tree/master/winPEAS}} is a post-exploitation tool developed to help security professionals, penetration testers and ethical hackers identify opportunities for escalation of privileges in Windows systems.
Our test results show that the default Windows Defender in Windows 11 Professional is able to thwart WinPEAS.
Similarly, commands seeking to access sensitive directories and files, such as the Security Accounts Manager (SAM) file and \texttt{tasklist}, were blocked, proving that endpoint defenses provided robust real-time security.

\textbf{Privilege Escalation Attempts:} \textbf{Mimikatz}~\footnote{\scriptsize \url{https://github.com/ParrotSec/mimikatz}}, a powerful post-exploitation tool that allows users to view and save authentication credentials such as Kerberos tickets, is used in privilege escalation tests for credential dumping.
These attempts are effectively mitigated by Microsoft Credential Guard, which prevented access to critical credential storage.
The usage also ensures that there is visibility of host-based actions and executables allowing us to monitor even down to what dynamic link libraries (DLLs) are called.

\textbf{{Access to ELK Stack}:} During the testing process, we identified misconfigurations during the scans where the ELK stack opens ephemeral ports and allows unrestricted access. An initial access to a workstation in the EPN led to the discovery of unintended access to an internally hosted ELK stack in \texttt{U1}, which is caused by the use of an earlier ELK version (e.g. Elasticsearch 7.17) for which security is not enabled automatically during the installation. 
In particular, the ELK stack does not require authentication, allowing unrestricted access to the Elasticsearch database and the Kibana dashboard.
The access effectively provides insight into the defensive monitoring capabilities.
The ELK stack is reinstalled fully with security features and confirms that access required authentication.

\input{figures/conn}
%
\input{figures/fig-conn-open-ports}
\input{figures/fig-failed-logins}
\subsection{Testing Detection and Alerts with C2 Malware}
As part of the validation process, we have also tested the \honeywin\ with 2 sets of malware on the ELK stack with the Elastic Defend setup.
One malware is a custom designed malware that provides a reverse shell via the \texttt{`netcat'} connection.
The malware provides a reverse shell connection, which allows an adversary to remotely control the endpoint through the command line interface (CLI).
We created this malware as a custom shellcode which connects to the adversary IP address and port while providing a CLI to interact with the endpoint upon execution.

Another malware was a modified Havoc malware.
Havoc~\footnote{\scriptsize \url{https://github.com/HavocFramework/Havoc}} is a post-exploitation and C2 framework designed for red-teaming and adversary simulation.
We created a Havoc agent with customized settings for stealth and evasive properties.
In this case, the agent runs on the endpoint and establishes a connection to the adversary's Havoc listener, enabling the adversary to remotely control the endpoint by sending commands.
Both malware were designed to be able to bypass Windows Defender, but intended to be detected by Elastic Defend.
Elastic Defend is able to detect both malware through their prebuilt Malware Detection Alerts rules.

%% file: figures/conn.tex
\begin{figure*}
    \centering
    \begin{subfigure}{\textwidth}
        \centering
        \includegraphics[width=\textwidth,keepaspectratio]{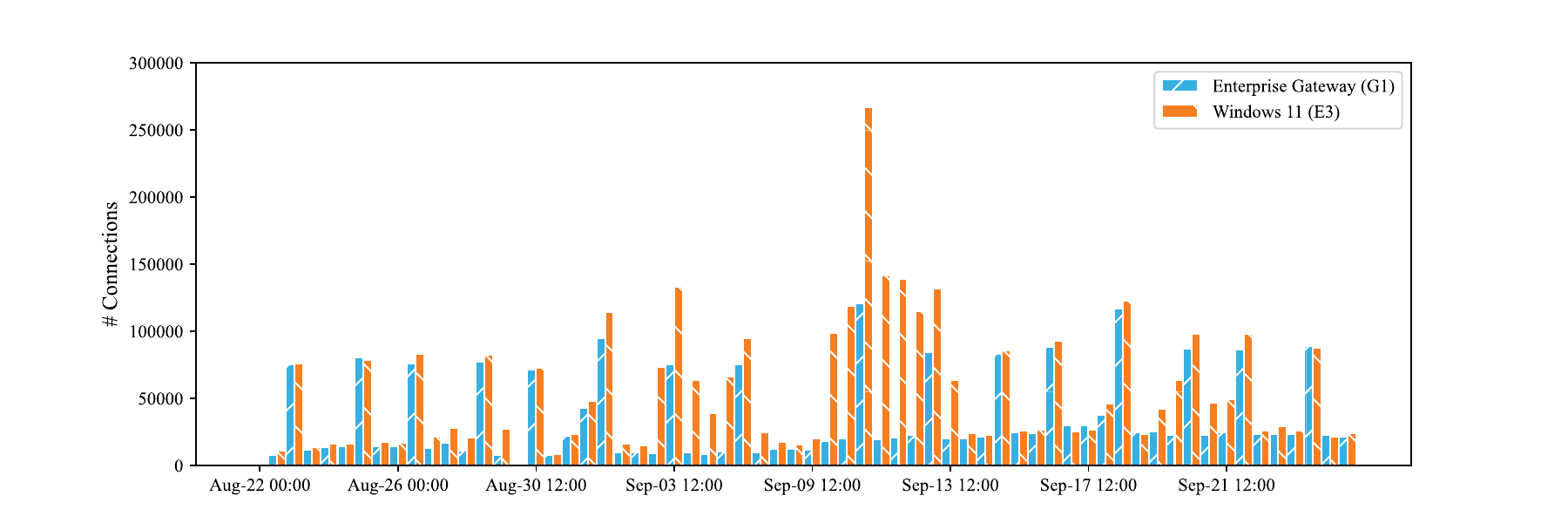}
        \caption{Incoming Connections Received by G1 and E3}
        \label{fig-conn-gateway-windows}
        \Description{}
    \end{subfigure}    
    \begin{subfigure}{0.47\textwidth}
        \centering
        \resizebox{0.95\linewidth}{!}{
        \begin{tikzpicture}       
            \pie[rotate = -90, text = pin]
            {38/United States, 27/Others, 6/United Kingdom, 7/The Netherlands, 9/Singapore, 13/Philippines}
        \end{tikzpicture}        
        }
        \caption{Geolocations of Incoming Connections Received by G1}
        \label{fig-geo-gateway}
        \Description{}
    \end{subfigure}
    \begin{subfigure}{0.47\textwidth}
        \centering
        \resizebox{0.80\linewidth}{!}{
        \begin{tikzpicture}       
            \pie[rotate = -90, text = pin]
            {43/Russia, 12/Others, 5/Singapore, 7/China, 14/Turkey, 19/United States}
            \end{tikzpicture}
        }
        \caption{Geolocations of Incoming Connections Received by E3}
        \label{fig-geo-windows}
        \Description{}
    \end{subfigure}
    \caption{Analysis of Incoming Connections Received by G1 and E3}
    \label{fig-conn}
\end{figure*}
%
%
%

%% file: figures/fig-conn-open-ports.tex
\begin{figure*}
    \centering    
    \begin{subfigure}{0.49\textwidth}
        \centering        
        \includegraphics[width=\textwidth,keepaspectratio]{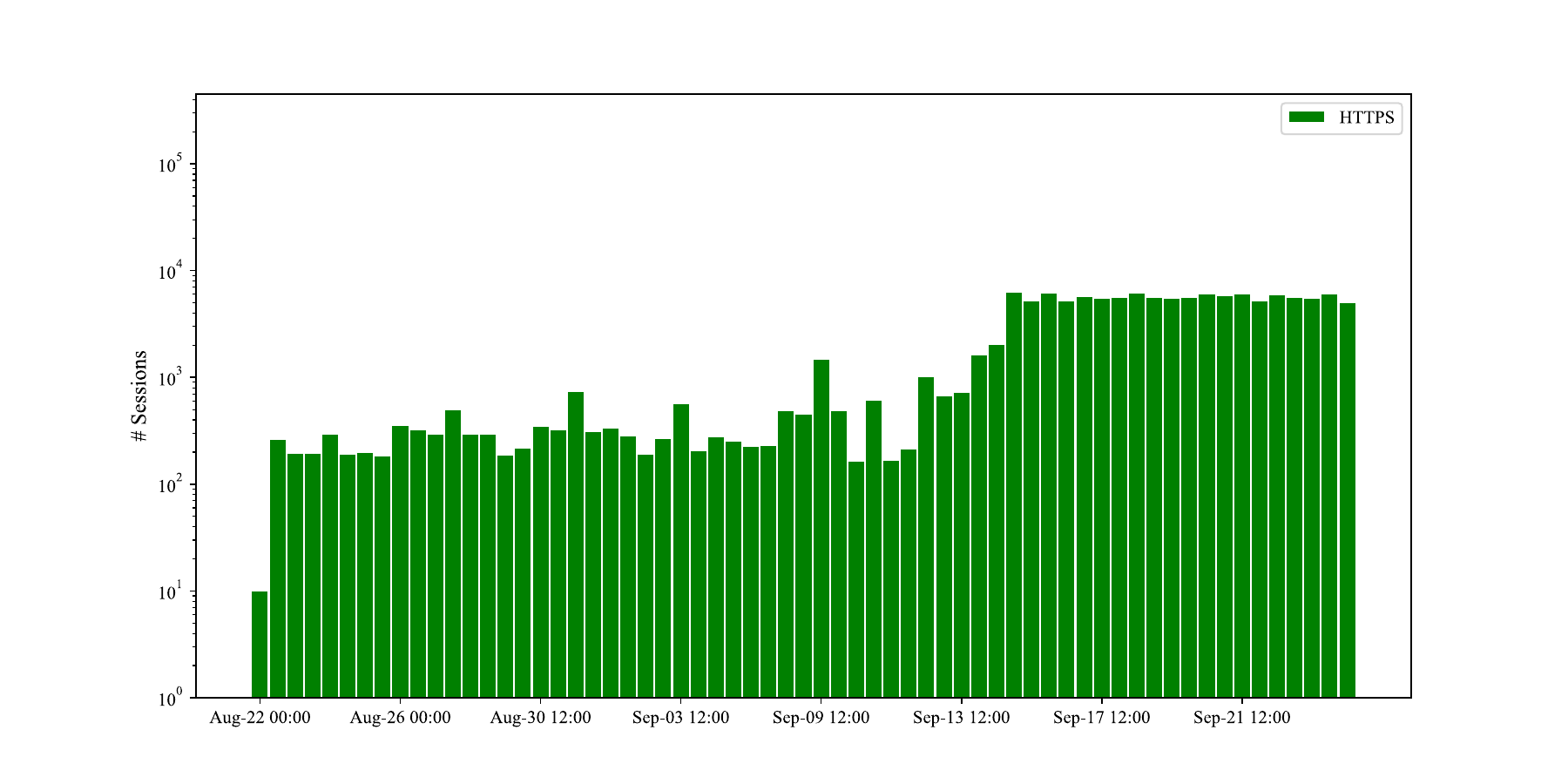}
        \caption{HTTPS Sessions of Enterprise Gateway (G1)}
        \label{fig-https-gateway}
        \Description{}
    \end{subfigure}
    \begin{subfigure}{0.49\textwidth}
        \centering        
        \includegraphics[width=\textwidth,keepaspectratio]{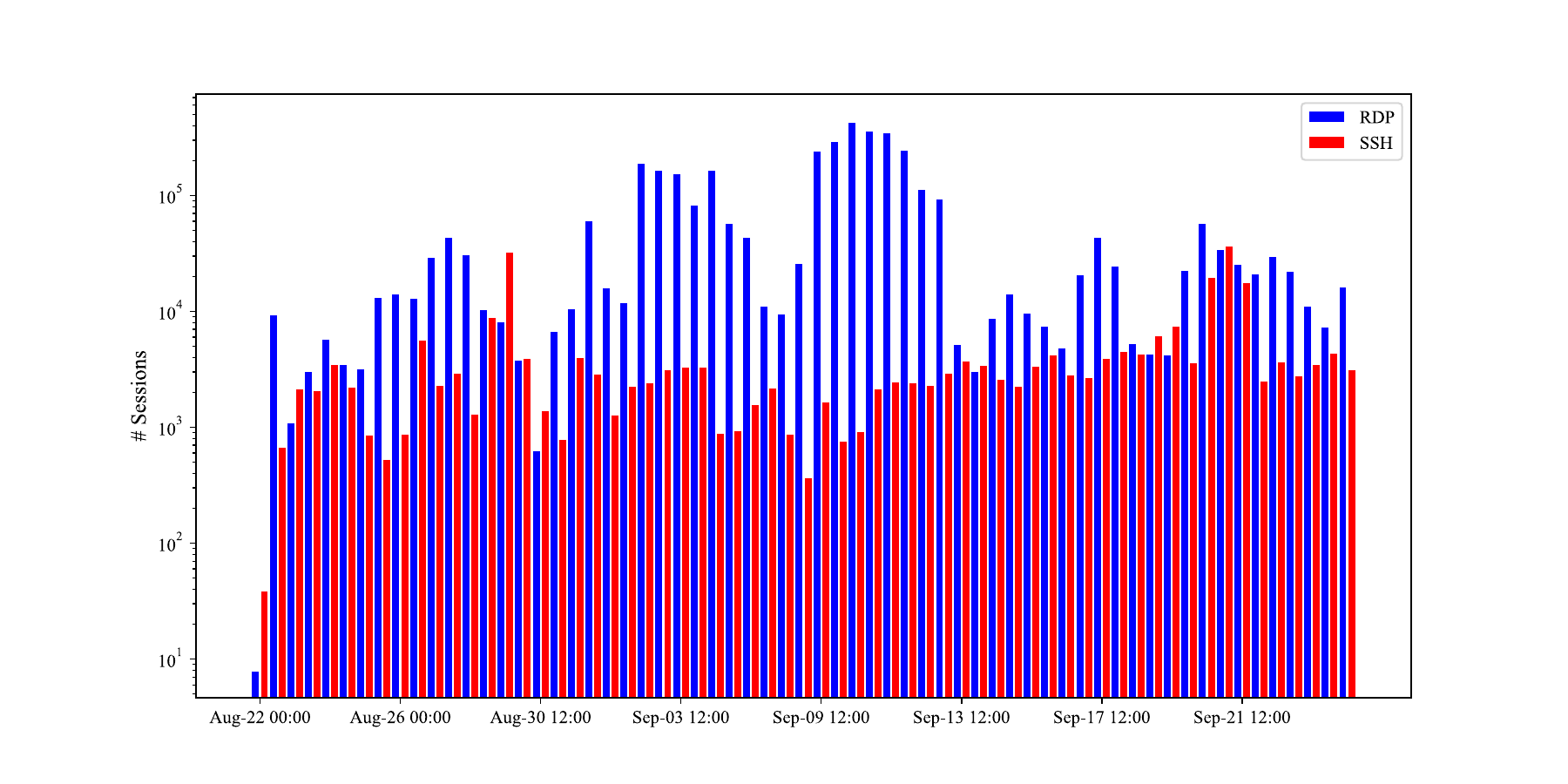}
        \caption{RDP \& SSH Sessions of Windows 11 (E3)}
        \label{fig-rdp-ssh-windows}
        \Description{}
    \end{subfigure}    
    \caption{Connections to Open Ports of G1 and E3}
    \label{fig-conn-open-ports}
\end{figure*}

%% file: figures/fig-failed-logins.tex
\begin{figure*}  
  \includegraphics[width=\textwidth,keepaspectratio]{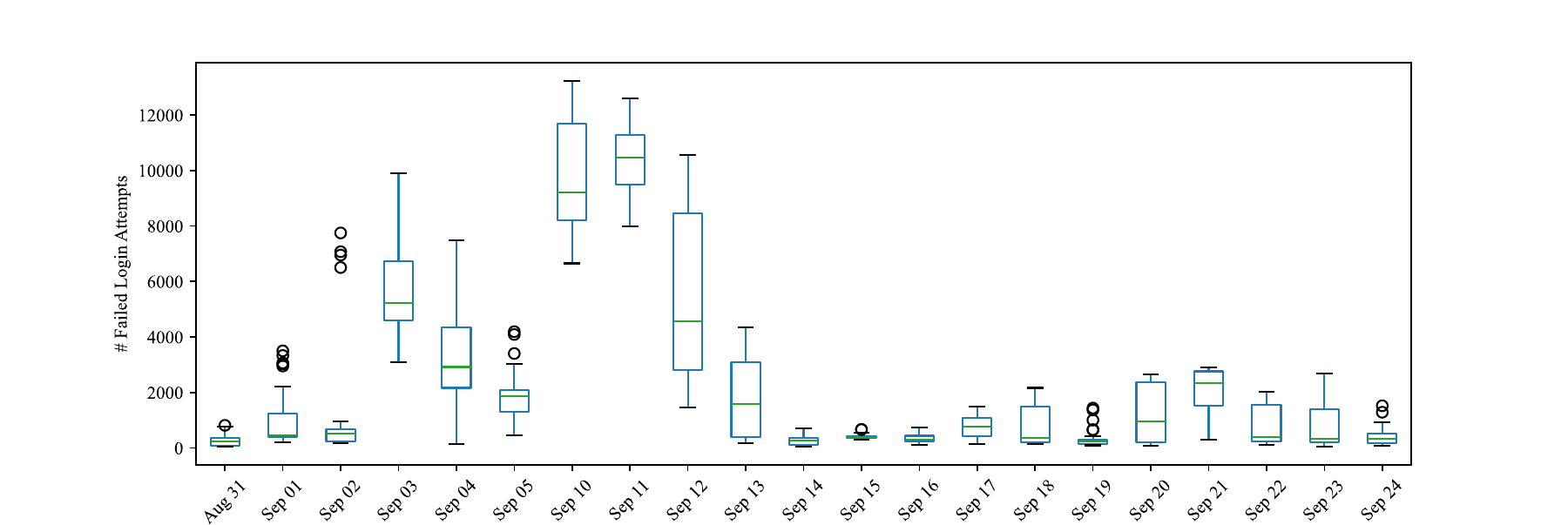}
  \caption{Boxplot of Failed Login Attempts}
  \label{fig-failed-logins}
  \Description{}
\end{figure*}

%% file: results.tex
\section{Experimental Results}
We have deployed the \honeywin\ system live in the wild for 34 days from 22 August 2024 to 25 September 2024.
This section provides our analysis on incoming and outgoing network traffic, failed login attempts, successful breaches via the exposed services, and the attacks initiated by the adversaries.

\subsection{Network Traffic Analysis} \label{subsec:network}
During this operating period, \honeywin\ received $\sim$5.79 million unsolicited connections, of which the gateway received 2.22 million connections and the Windows endpoint (\texttt{E3}) received 3.57 million connections.
In general, the Windows endpoint (\texttt{E3}) received 60\% more connections than the gateway (\texttt{G1)}.
Figure~\ref{fig-conn} shows the distribution of the connections received by \texttt{G1} and \texttt{E3} over a 12-hour interval for 34 days.
\texttt{G1} received more connections than \texttt{E3} for 8 intervals only.

Figures~\ref{fig-geo-gateway} and \ref{fig-geo-windows} show geolocations based on IP address lookup of incoming connections received by \texttt{G1} and \texttt{E3}, respectively.
For the gateway (\texttt{G1}), 38\% and 13\% of the connections are initiated from IP addresses belonging to United States and Philippines.
For the Windows endpoint (\texttt{E3}), 43\% and 19\% of the connections are from IP addresses belonging to Russia and the United States.
It should be noted that the identification of geolocations from IP address lookup is only indicative since the attackers may have spoofed their true IP addresses.

Figures~\ref{fig-https-gateway} and \ref{fig-rdp-ssh-windows} show connections to open port of the gateway (\texttt{G1}) and Windows engpoint (\texttt{E3}) over 12-hour intervals for 34 days.
The vertical axis in both figures shows the number of sessions in logarithmic scale.
The gateway (\texttt{G1}) receives more than 2,123 HTTPS sessions per 12 hours on average.
On the other hand, the Windows endpoint (\texttt{E3}) receives more than 31,972 RDP and SSH sessions combined every 12 hours on average.
This indicates that the Windows endpoint (\texttt{E3}) is more than 15 times active compared to the gateway (\texttt{G1}).
Furthermore, the Windows endpoint (\texttt{E3}) receives 59,591 RDP and 4,263 SSH sessions per 12 hours on average.
The attacker actively exploits RDP nearly 14 times more than SSH.
%
%
\subsection{Failed Login Attempts}
As described in Section~\ref{subsec:implementation}, \honeywin is capable of capturing failed login attempts for the Windows endpoint (\texttt{E3}).
Figure~\ref{fig-failed-logins} shows a boxplot of failed login attempts for \texttt{E3}.
The vertical axis shows the number of failed login attempts per hour for each day on the horizontal axis.
The data range varies over the \honeywin\ operating period.
We observe significant outliers in certain days (e.g., Sep 01, 02 and 05).
Also, significantly higher first-quartile (Q1) and median (Q2) values (e.g., Sep 10 and 11).
This strongly indicates active brute-forcing attempts by the attackers during certain hours of the day or throughout certain days.
%
\subsection{Successful Logins}
During the operating period, the \honeywin\ detected 5 successful logins via RDP and 354 successful logins via SSH.
We have analyzed all successful log-ins and developed an attack attribution method which correlates the incoming network traffic, host logs, and outgoing traffic holistically.
For 4 out of 5 successful logins via RDP, we find that these logins are short-lived and do not see any activities performed by the attackers after the successful logins.
Section~\ref{subsec:tokens} provides our analysis on the remaining successful login session during which the attacker interacted with our deceptive tokens.
As described in Section~\ref{subsec:network}, although the attacker actively exploits RDP nearly 14 times more than SSH, our analysis shows that a stealthy attack was launched through successful SSH logins.
We provide detailed analysis of this attack in Section~\ref{subsec:ssh}.

However, no successful login alerts are generated for the gateway (\texttt{G1}) and two Windows endpoints (\texttt{E1} and \texttt{E2}).
We conjecture that strong password setting of the gateway (\texttt{G1}) prevented the breach, although there are sustained brute-force attempts.
For the case of \texttt{E1} and \texttt{E2}, the endpoints are on the EPN with NAT connections to the \texttt{G1}.
This imposes an additional layer of difficulty that requires one to bypass the gateway {\texttt{G1}} to reach both endpoints.
%
%
\input{figures/fig-rdp}
\subsection{Analysis on Deceptive Token Interaction} \label{subsec:tokens}
Analysis of system logs and network traffic reveals that a successful RDP session was established using one of the standard user accounts of \texttt{E3}.
Correlating process creation logs with network traffic, we are able to trace the attacker's activities and interactions with our deceptive tokens.

Based on the timestamp of the successful RDP login alert notification, we identify the RDP login event and correlate it with the process creation timestamps.
Typically, when a user logs in via RDP, the system spawns key processes such as \texttt{`smss.exe'}, \texttt{`winlogon.exe'}, and \texttt{`userinit.exe'}.
The sequential execution of these processes confirmed a successful login as illustrated in Figure~\ref{fig-rdp}.

Further analysis of the process creation sequence reveals that \texttt{`powershell.exe'} was launched by \texttt{`explorer.exe'}, indicating post-login activity.
Notably, there is a two-minute interval between the login and the execution of \texttt{powershell.exe}.
This suggests that the reconnaissance is performed manually by a user rather than through an automated script, reinforcing the likelihood of an active attacker on the keyboard.

A key observation is that \texttt{`powershell.exe'} subsequently executed our deceptive token, \texttt{`systeminfo.exe'}.
This confirms that the attacker has been engaging with the deceptive environment, believing it to be a legitimate endpoint.
\input{figures/fig_ssh}
\input{algorithms/alg-attack-attribution}
\input{figures/incoming-ssh}
\input{tables/tab-smtp-creds}
\subsection{Stealthy SMTP Brute-Force Bot Attack via Successful SSH Logins}
\label{subsec:ssh}
We received a successful SSH login alert at 08:40 PM on 18 September 2024 (GMT), followed by two hourly successful logins until 04:50 PM, 23 September 2024.
Our initial analysis on these successful logins does find any sign of attack.
Subsequently, the successful logins become every two minutes.
Although we did not find significant host activities up to this time, the SPAN port is capturing 100 MB of network traffic every 7 minutes.
Further investigation shows reverse SSH tunneling after the initial successful SSH login spawning child SSH processes.
In this case, SMTP outgoing traffic could be observed from child SSH processes.
However, the ELK is not able to log the SMTP payload due to reverse SSH tunneling crippling host-based detection and insights.
Since all outgoing traffic originating from the \honeywin\ is captured via the SPAN port, we are able identify the SMTP payloads from the network traffic.
The adversary was performing an attack on global SMTP domains via successful SSH logins recruiting the endpoint (\texttt{E3}) into a botnet.
Figure~\ref{fig-ssh} shows our analysis of successful SSH Logins followed by SMTP brute-
force bot attack.

It is evident that host-based logging alone does not suffice and shed light on the advantage of \honeywin\ as it captures incoming network traffic, host logs and outgoing network traffic, all together thereby facilitating a holistic detection capability.
We have developed an attack attribution algorithm to establish an end-to-end correlation of host logs and network traffic (see Algorithm~\ref{alg:attackattributionssh}).

We are able to trace the attacker’s activities and determine that SMTP brute force bot attacks have been carried out on Internet-facing SMTP servers.
Figure~\ref{fig-geo-in-ssh} shows the geolocations of 354 successful SSH logins based on IP address lookup
of incoming network traffic connections, noting that attackers could have spoofed their true IP addresses.
Similarly, Figure~\ref{fig-geo-out-ssh} shows geolocations of outgoing network traffic initiated by the attacker.
In this case, geolocations may not have spoofed since these are the attacker's targets.
In fact, the outgoing traffic indicates geolocations of 180 countries while the United States taking up 19\% followed by France and Germany with 10\% each, Poland (8\%), Japan (6\%) and The Netherlands (4) where the remaining 43\% belongs to 174 countries.
In terms of outgoing port distribution, port 25 (Standard) takes up 74\%, 587 (Default) and 465 (TLS) are 22\% and 4\% respectively.

Our analysis shows that the attacker used 151,179 credentials.
Table~\ref{tab:t15usernames} provides the top 15 SMTP usernames, while Table~\ref{tab:t15passwords} and 
\ref{tab:t15usrpwd} list the top 15 passwords and username-password pairs used by the attacker.
The brute-force attempts harvested 1250 successful SMTP credentials.
The attacker uses the email addresses in Table~\ref{tab-smtp-emails} to record the successful credentials upon acceptance by the SMTP server.
The successful SMTP credentials do not appear to be randomly generated, rather they are specific to certain users.
These credentials likely originated from a data breach, and we choose not to reveal them in this work.
%
\input{tables/tab-smtp-emails}

%% file: figures/fig-rdp.tex
\begin{figure}[h]  
  \includegraphics[width=0.95\linewidth,keepaspectratio]{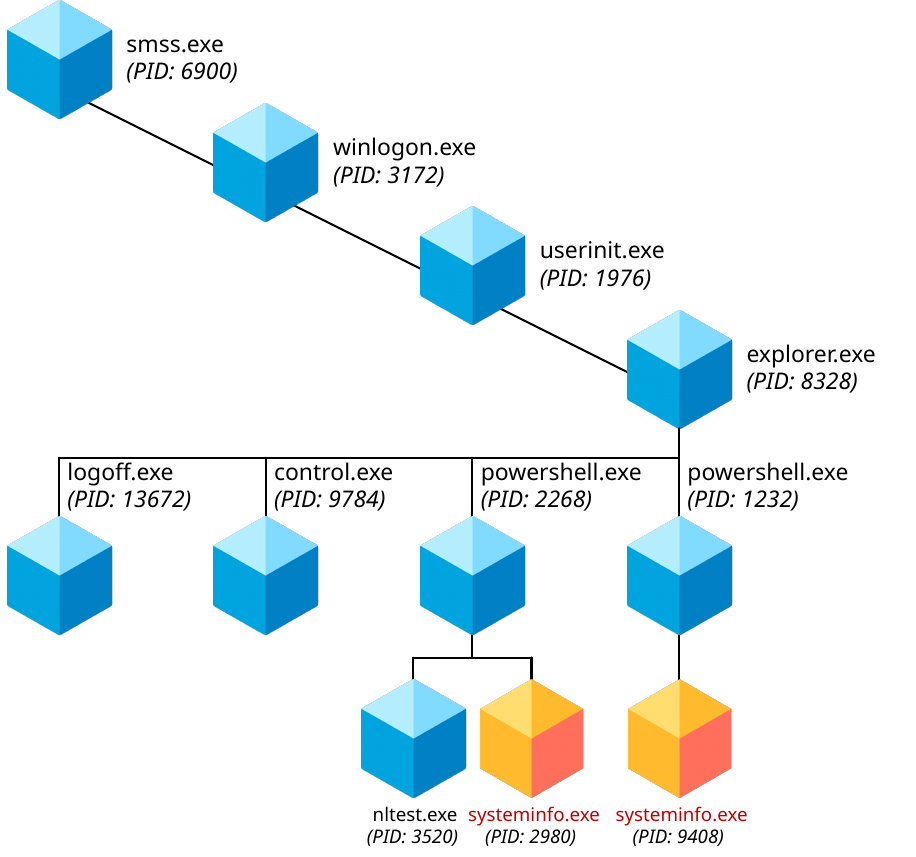}
  \caption{Analysis of a Successful RDP Login \& Deceptive Token Interaction by the Adversary}
  \label{fig-rdp}
  \Description{}
\end{figure}

%% file: figures/fig_ssh.tex
\begin{figure}  
  \includegraphics[width=0.95\linewidth,keepaspectratio]{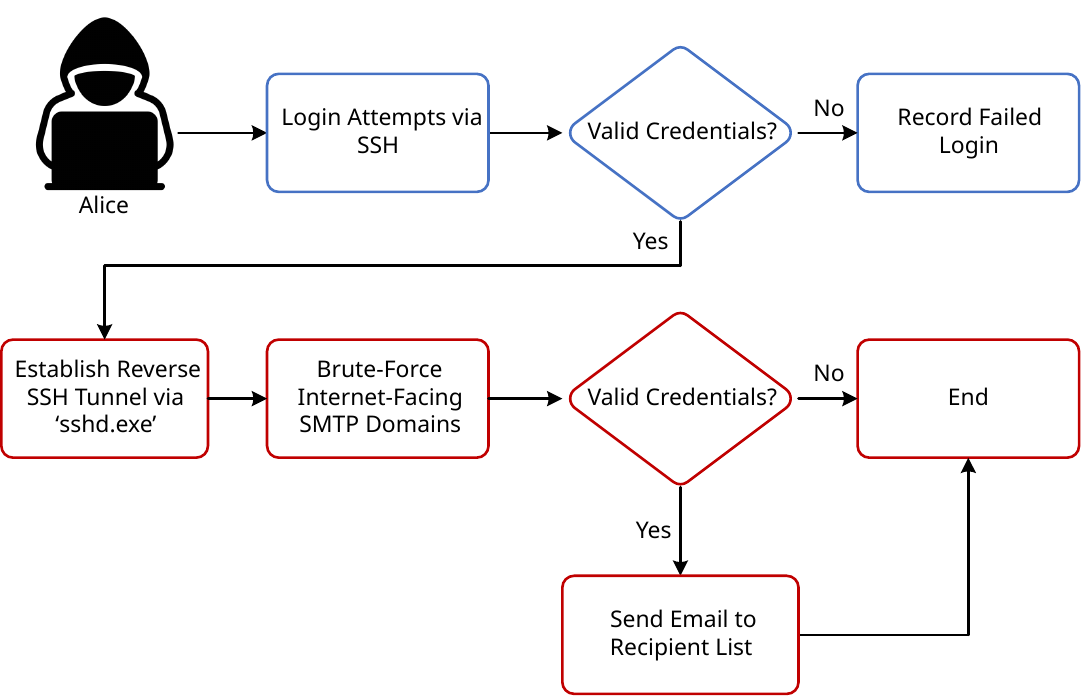}
  \caption{Analysis of Successful SSH Logins \& SMTP Brute-Force Bot Attack}
  \label{fig-ssh}
  \Description{}
\end{figure}

%% file: algorithms/alg-attack-attribution.tex
\begin{algorithm}[t]
\caption{Attack-Attribution-SSH}\label{alg:attackattributionssh}
\begin{algorithmic}[1]
\Require
Successful SSH login alerts $\textbf{A}$, Incoming traffic PCAPs $\textbf{N}$, SPAN port PCAPs $\textbf{S}$, Elasticsearch database $\textbf{E}$
\Ensure
Incoming IPs $\textbf{I}$, Outgoing IPs \& Ports $\textbf{O}$, Payload: $\textbf{P}$
\State Get timestamps $\textbf{T}$ from alerts $\textbf{A}$
\State Construct ElasticSearch query $\textbf{Q}$ for $\textbf{T}$
\State Construct TShark query $\textbf{F}$ for $\textbf{T}$
\State $I$ = \Call{getIncomingAttackerIp}{$\textbf{N}$, $\textbf{F}$}
\State $O$ = \Call{getOutgoingDataFromElastic}{$\textbf{E}$, $\textbf{Q}$}
\State Construct TShark query $\textbf{G}$ to extract SMTP payload from $\textbf{O}$
\State $P$ = \Call{getOutgoingSmtpPayload}{$\textbf{S}$, $\textbf{G}$}
\State \Return $\textbf{I}$, $\textbf{O}$ and $\textbf{P}$
\Statex
\Procedure{getIncomingAttackerIp}{$\textbf{N}$, $\textbf{F}$} \label{alg:getIncomingAttackerIp}
\State Get incoming attacker IPs $\textbf{I}$ from PCAP $\textbf{N}$ with query $\textbf{F}$
\State \Return $\textbf{I}$
\EndProcedure
\Statex
\Procedure{getOutgoingDataFromElastic}{$\textbf{E}$, $\textbf{Q}$} \label{alg:getOutgoingDataFromElastic}
\State Initialize Elasticsearch connection
\State Get outgoing IPs \& ports $\textbf{O}$ from Elasticsearch $\textbf{E}$ with $\textbf{Q}$
\State \Return $\textbf{O}$
\EndProcedure
\Statex
\Procedure{getOutgoingSmtpPayload}{$\textbf{S}$, $\textbf{G}$} \label{alg:getOutgoingSmtpPayload}
\State Get outgoing SMTP data (\texttt{username}, \texttt{password}, \texttt{mail to}, \texttt{mail from}, etc) $\textbf{P}$ from $\textbf{S}$ with query $\textbf{G}$
\State return $\textbf{P}$
\EndProcedure
\end{algorithmic}
\end{algorithm}

%% file: figures/incoming-ssh.tex
\begin{figure*}
    \centering
    \begin{subfigure}{0.47\textwidth}
        \centering
        \resizebox{0.80\linewidth}{!}{
        \begin{tikzpicture}       
            \pie[rotate = -90, text = pin]
            {40/Germany, 5/Others, 3/Canada, 4/China, 7/The Netherlands, 15/Bulgaria, 26/France}
        \end{tikzpicture}
        }
        \caption{Geolocations of Incoming Connections for Successful SSH Logins}
        \label{fig-geo-in-ssh}
        \Description{}
    \end{subfigure}
    \begin{subfigure}{0.47\textwidth}
        \centering
        \resizebox{0.80\linewidth}{!}{
        \begin{tikzpicture}       
            \pie[rotate = -90, text = pin]
            {43/Others, 4/The Netherlands, 6/Japan, 8/Poland, 10/Germany, 10/France, 19/United States}
            \end{tikzpicture}
        }
        \caption{Geolocations of Outgoing Connections for Successful SSH Logins}        
        \label{fig-geo-out-ssh}
        \Description{}
    \end{subfigure}
    \caption{Analysis of Incoming and Outgoing Connections for Successful SSH Logins}
    \label{fig-in-ssh}
\end{figure*}
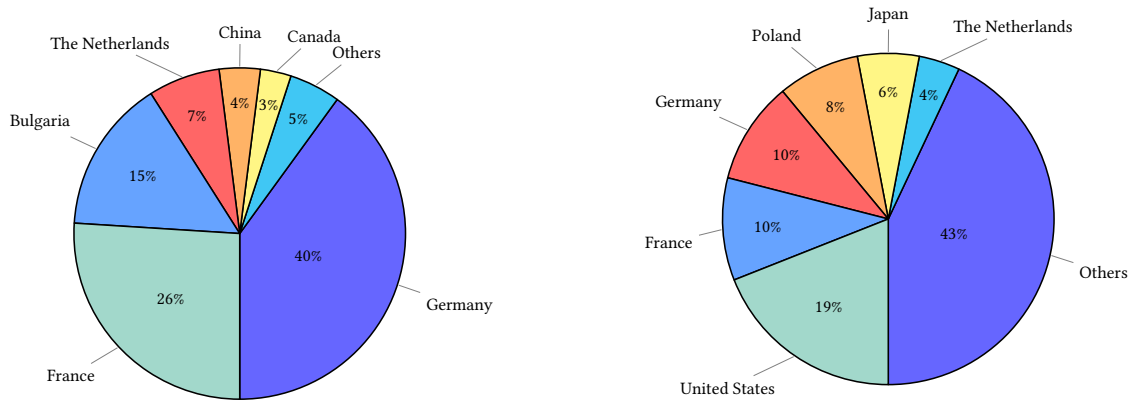

%% file: tables/tab-smtp-creds.tex
%
\begin{table*}
    \centering
    \label{tab:smtp-creds}
    \begin{minipage}[b]{.30\linewidth}
        \centering
        \caption{Top 15 SMTP Usernames}
        \begin{tabular}{l|r}
        \toprule \toprule
        Username & \# Occurrence \\
        \midrule
        \texttt{info 	   } & \texttt{2529} \\
        \texttt{contact    } & \texttt{ 562} \\
        \texttt{no-replay  } & \texttt{ 483} \\
        \texttt{admin      } & \texttt{ 465} \\
        \texttt{office     } & \texttt{ 458} \\
        \texttt{inf        } & \texttt{ 389} \\
        \texttt{marketing  } & \texttt{ 326} \\
        \texttt{mail       } & \texttt{ 324} \\
        \texttt{support    } & \texttt{ 315} \\
        \texttt{webmaster  } & \texttt{ 314} \\
        \texttt{hello      } & \texttt{ 248} \\
        \texttt{noreply    } & \texttt{ 246} \\
        \texttt{root       } & \texttt{ 223} \\
        \texttt{test       } & \texttt{ 215} \\
        \texttt{postmaster } & \texttt{ 214} \\
        \bottomrule \bottomrule
        \end{tabular} 
        \label{tab:t15usernames}
    \end{minipage}
    \begin{minipage}[b]{.30\linewidth}
        \centering
        \caption{Top 15 SMTP Passwords}
        \begin{tabular}{l|r}
        \toprule \toprule
        Password & \# Occurrence \\
        \midrule
        \texttt{co23     } & \texttt{587} \\
        \texttt{in23     } & \texttt{398} \\
        \texttt{ma23     } & \texttt{380} \\
        \texttt{123456   } & \texttt{241} \\
        \texttt{q1w2e3r4 } & \texttt{194} \\
        \texttt{ne23     } & \texttt{184} \\
        \texttt{123654   } & \texttt{180} \\
        \texttt{test123  } & \texttt{177} \\
        \texttt{11223344 } & \texttt{175} \\
        \texttt{P@ssw0rd } & \texttt{175} \\
        \texttt{1234     } & \texttt{174} \\
        \texttt{Abcd1234 } & \texttt{172} \\
        \texttt{1q2w3e4r } & \texttt{170} \\
        \texttt{11111111 } & \texttt{166} \\
        \texttt{sa23     } & \texttt{164} \\
        \bottomrule \bottomrule
        \end{tabular}        
        \label{tab:t15passwords}
    \end{minipage}   
    \begin{minipage}[b]{.38\linewidth}
        \centering
        \caption{Top 15 Username \& Password Pairs}
        \begin{tabular}{l|r|r}
        \toprule \toprule
        Username & Password & \# Occurrence \\
        \midrule
        \texttt{info}	   & \texttt{in23}	       & \texttt{128} \\
        \texttt{info}	   & \texttt{inf23}	       & \texttt{ 39} \\
        \texttt{marketing} & \texttt{marketing@1}  & \texttt{ 31} \\
        \texttt{hello}     & \texttt{Hello@15}	   & \texttt{ 28} \\
        \texttt{no-replay} & \texttt{noreplay@1}   & \texttt{ 27} \\
        \texttt{test}	   & \texttt{test@1}	   & \texttt{ 27} \\
        \texttt{webmaster} & \texttt{Webmaster@15} & \texttt{ 26} \\
        \texttt{contact}   & \texttt{Contact@1}	   & \texttt{ 25} \\
        \texttt{contact}   & \texttt{Contact@15}   & \texttt{ 24} \\
        \texttt{root}	   & \texttt{Root@1}	   & \texttt{ 24} \\
        \texttt{contact}   & \texttt{contact@15}   & \texttt{ 24} \\
        \texttt{contact}   & \texttt{Contact2021}  & \texttt{ 23} \\
        \texttt{office}	   & \texttt{office@1}	   & \texttt{ 23} \\
        \texttt{root}	   & \texttt{root@1}	   & \texttt{ 22} \\
        \texttt{root}	   & \texttt{root@15}      & \texttt{ 22} \\
        \bottomrule \bottomrule
        \end{tabular}        
        \label{tab:t15usrpwd}
    \end{minipage}        
\end{table*}

%% file: tables/tab-smtp-emails.tex
\begin{table}[!htbp]
    \centering
    \caption{Attacker Email Addresses}
    \label{tab-smtp-emails}
    \begin{tabular}{l|r}
    \toprule \toprule
    Email Address & \# Occurrence \\
    \midrule
    \texttt{toron@imobust.com           } & \texttt{172} \\
    \texttt{no-reply-1@mx-test-serv.org } & \texttt{155} \\
    \texttt{no-reply10@mx-test-serv.org } & \texttt{147} \\
    \texttt{c2@mail-master.org          } & \texttt{136} \\
    \texttt{bt@mail-master.org          } & \texttt{126} \\
    \texttt{c4@mail-master.org          } & \texttt{101} \\
    \texttt{c3@mail-master.org          } & \texttt{98} \\
    \texttt{c1@mail-master.org          } & \texttt{86} \\
    \texttt{check@bewareofdogs.xyz      } & \texttt{85} \\
    \texttt{no-reply-2@mx-test-serv.org } & \texttt{80} \\
    \texttt{mail2@glob22glo1.su         } & \texttt{41} \\
    \texttt{no-reply12@mx-test-serv.org } & \texttt{26} \\
    \bottomrule \bottomrule
    \end{tabular}        
\end{table}

%% file: discussion.tex
\section{Discussion}
\label{sec:discussion}
%
%
This section discusses our insights from the design and implementation of the \honeywin\ system and the experimental results.
Firstly, three endpoints (i.e., \texttt{E1}, \texttt{E2} and \texttt{E3}) deployed with the \honeywin\ system are not virtual machines.
They are rather real devices (e.g., Mini PC) running a fresh out-of-the-box (OOB) Windows 11 Professional with latest updates and automatic updates enabled.
The deceptive tokens and bait files are installed on top of the OOB.
As provided in Table~\ref{tab-deceptive-tokens}, the deceptive tokens are installed system-wide for all users.
However, bait files are placed for certain standard user accounts only recalling that \texttt{<E2>} and \texttt{<E3>} are placeholders for standard users.
There are no bait files for the standard user account breached for the E3 endpoint during the successful RDP session where the adversary interacted with the deceptive token.
This explains the fact that the adversary is not able to interact with the bait files in this case and highlights the need for a scheme to distribute the bait files and the deceptive tokens automatically across all the endpoints.

Successful SSH log-ins for the SMTP brute-force bot attack occurred every two hours initially, indicating that it was likely a scripted attack.
The use of reverse SSH tunnels reveals the adversary's intention to avoid host-based detection and maintain stealth as long as possible.
Successful SSH logins may have been classified as false positives, and the attack could have been missed in the absence of outgoing network traffic capture.

While this work focuses on the HoneyWin deployment in an enterprise environment, we could extend \honeywin\ to an OT environment with HMI and SCADA endpoints, potentially making it more enticing for the adversary.

%% file: related_work.tex
\section{Related Work}
\label{sec:related-work}
Honeypots are security resources whose value lies in their ability to be probed, attacked and compromised.
In general, there are two types of honeypots: first, low-interaction honeypots, which present simulated or emulated services/environments to attackers, and second, high-interaction honeypots, which show real systems to attackers.
The characteristics of honeypots are that they are deceptive, discoverable, interactive, and monitored.

Several prior efforts have leveraged honeypot systems as a proactive mechanism for network defense, recognizing their potential to gather in-depth threat intelligence and lure attackers away from real assets.
In \cite{Han2018Deception}, the authors highlighted the value of honeypots to test adversaries’ behaviors in controlled yet realistic environments, while other works have focused on enhancing honeypot scalability and adaptability in complex networks.
As Grimes described, every corporate entity should be running honeypots if they are interested in the earliest warning possible of a successful hack or malware infiltration ~\cite{grimes2017hacking}.

Various honeypot systems such as honeypots for database systems, web applications, services, ICS/SCADA, etc., have been proposed and developed by industry and academia~\cite{awesomehoneypots2017online}.
Notable implementations include Cowrie~\cite{cowire:online}, Glutton~\cite{glutton:online}, OpenCanary~\cite{opencanary:online}, Conpot~\cite{conpot:online}, T-Pot~\cite{tpot:online}, AIDE~\cite{aide:online}, ICSNet~\cite{icsnet}, SIPHON~\cite{guarnizo2017siphon} etc.
As mentioned, Windows OS are ubiquitous in enterprise IT and (OT environments.
Since attacks targeting Windows-based systems have been on the rise in recent years, we provide our review with a focus on related state-of-the-art Windows-based honeypot implementations.

KFSensor, launched in 2003, is one of the first Windows-based honeypot intrusion detection systems to attract and detect hackers and worms by simulating vulnerable system services and trojans~\cite{kfsensor:online}.
It is available in Professional, Enterprise, and Educational editions.
KFSensor could be configured using scenarios, for example, to listen on all TCP and UDP ports, to simulate a MySQL database server or an IIS Web server.
Since it simulates vulnerable services, it is likely that attackers will detect that they are interacting with a honeypot rather than a real system.
Moreover, KFSensor is a proprietary software that cannot be modified to tailor specific research purposes.

Wang et al. proposed the Strider HoneyMonkey Exploit Detection System, which uses a network of monkey programs running on virtual machines with different patch levels to hunt for websites that exploit browser vulnerabilities~\cite{honey:monkeys}.
The authors identified 752 unique URLs operated by 287 websites that can successfully exploit unpatched Windows XP machines.

Microsoft has been developing a honeypot sensor network since 2018~\cite{msdeception:online}.
The honeypot framework, written in C\#, allows security researchers to quickly deploy various types of exploit handlers, from simple HTTP handlers to complex protocols such as SSH and VNC.
In 2021, a dangling subdomain, \texttt{code.microsoft.com}, was used temporarily to host a malware C2 service.
Instead of removing the subdomain, Microsoft redirected it to the honeypot sensor network till 26 April 2024.
The data and findings have been published in~\cite{passino2024nested} and are crucial for understanding the 0day and nDay~\footnote{\scriptsize nDay vulnerability is a security flaw that has been disclosed and has an official patch available. However, organizations may still be unpatched and vulnerable.} ecosystem.
The framework itself is proprietary, and Microsoft have put in substantial engineering effort into this.
Attackers can communicate with over 30 different protocols/services.

HopLab~\cite{hoplab:online}, proposed by the Laboratoire de Haute Sécurité (LHS) in Rennes, deploys high-interaction honeypot systems to attract and analyze malicious activities.
HopLab is designed for rapid and flexible deployment of honeypots to respond to emerging vulnerabilities such as the Log4j. 
It can swiftly set up honeypots that emulate these specific weaknesses across various environments, including both Windows and Linux OS enabling timely analysis of new threats as they arise.
Although direct comparison with existing state-of-the-are works has been very limiting, \honeywin complements and contributes to ongoing research on high-interaction Windows honeypots, offering an additional layer of monitoring, while also illuminating best practices for honeypot configuration and risk mitigation.

In addition, recent studies further highlight honeytoken deployment and deceptive services as means of detecting novel or stealthy attacks.
Han et al.~\cite{Han2018Deception} present a comprehensive examination of deception techniques within computer security, classifying solutions such as honeypots, honeytokens and obfuscation according to various layers (network, system, application, and data) and core objectives (prevention, detection or mitigation of attacks).
The survey emphasizes the challenges of generating realistic decoys, optimally placing them, and continuously updating deception elements to avoid attacker evasion.
Although the findings show that deception can effectively complement traditional defenses, such as intrusion detection systems, critical gaps remain, including the need for reproducible experiments, robust evaluation methodologies, and reliable ways to seamlessly integrate deception into broader security strategies.
\honeywin\ aims to resolve this issue with a configurable standardized set of deceptive tokens to ensure reproducible results.
%

%% file: conclusions.tex
\section{Conclusions} \label{sec:conclusions}
We have designed and implemented a high-interaction Windows honeypot system that mimics an enterprise IT environment, namely, \honeywin\ in this paper.
By deploying the \honeywin\, with three Windows endpoints and an enterprise grade gateway, live in the wild for 34 days it receives more than 5.79 million unsolicited connections, 1.24 million login attempts, 5 and 354 successful logins via RDP and SSH sessions respectively.
In addition, the adversary interacted with the deceptive token in one of the RDP sessions and exploited the public-facing Windows endpoint to initiate the SMTP brute-force bot attack.
The adversary successfully harvested 1,250 SMTP credentials after attempting 151,179 credentials during the attack.
The results indicate that the \honeywin\ system is enticing for adversaries to probe, compromise, and launch attacks.
Moreover, the \honeywin\ enables us to receive real-time intrusion alerts and attribute attacks via the comprehensive network traffic capturing and host logging capabilities.
While we focus the \honeywin\ deployment in an enterprise environment, we plan to extend this to an OT environment as part of our future work.
Ultimately, \honeywin\ could be scaled and harness the recent development in artificial intelligence technology to realize a robust security posture that detects, analyzes, and mitigates threats effectively.